# Random Processes with High Variance Produce Scale Free Networks

*Josh Johnston and Tim Andersen*

## 1 Abstract

Real-world networks tend to be scale free, having heavy-tailed degree distributions with more hubs than predicted by classical random graph generation methods. Preferential attachment and growth are the most commonly accepted mechanisms leading to these networks and are incorporated in the Barabási-Albert (BA) model [1]. We provide an alternative model using a randomly stopped linking process inspired by a generalized Central Limit Theorem (CLT) for geometric distributions with widely varying parameters. The common characteristic of both the BA model and our randomly stopped linking model is the mixture of widely varying geometric distributions, suggesting the critical characteristic of scale free networks is high variance, not growth or preferential attachment. The limitation of classical random graph models is low variance in parameters, while scale free networks are the natural, expected result of real-world variance.

## 2 Introduction and Background

## 2.1 Barabási Albert Model

The BA model states that degree – the number of links per node – is distributed in a real network as a power law due to growth and preferential attachment [2], [3]. The BA model is claimed for many complex systems that lack an obvious real-world mechanism for preferential attachment and its applicability in network science is debated [4]. But despite many papers identifying scale free networks in the real-world, few have looked for alternatives to preferential attachment when explaining how these networks are formed [5], [6].

Starting with an initial graph, at each time step a node is added to the network which links to *m* existing nodes (growth). These *m* links accrue to existing nodes with probability proportional to the degree of those nodes (preferential attachment). Nodes with higher degree are more likely to receive links from new nodes according to a linear function. The probability *Π(k)* that a link of the new node *t* connects to node *i* depends on the degree $k_i$ as

$$\Pi(k_i) = \frac{k_i}{\sum_{j=0}^{n} k_j}$$

Equation 1

This leads to the degree distribution:

$$p_{BA}(k) = \frac{2m(m+1)}{k(k+1)(k+2)} \text{ for } k \geq m$$

Equation 2

Where $p_{BA}(k)$ is the probability a node has degree *k* according to the BA model[7]. When *k* grows large, Equation 2 becomes a power law with *γ=2* because $p_{BA}(k)$ -> $k^{-3}$.

The World Wide Web (WWW) network is a directed graph where hyperlinks (links) point from one webpage (node) to another. The BA model states that newly added webpages will add a set number of hyperlinks, *m*, that preferentially point to existing webpages with a lot of links (Google, Wikipedia, etc). The model isn't intended to capture all complexity of a real system like the WWW, but is the simplest set of assumptions that leads to scale free networks caused by node growth and preferential attachment.

The BA model suggests an irresistible first-mover advantage that has informed growth-at-all-cost strategies in several industries. Growth in the BA model means links are formed when new nodes join the network, so older nodes have more chances to collect links. Preferential attachment further disadvantages late arrivals because links form with probability proportional to the degree of the existing nodes. Many complex systems show latecomers overcoming this initial advantage, including Google's search engine and Facebook's social network platform displacing well-established players. This required tempering the BA model with the concept of intrinsic fitness [8]. Incorporating fitness in the BA model does not affect the mechanisms leading to power law degree distributions, which remain preferential attachment and growth.

Previous research demonstrates fitness alone is sufficient to generate scale free networks in special cases, controlling linking with a function of the fitnesses of the two vertices involved that is selected appropriately for the particular fitness distribution, requiring symmetry between the two linking nodes[9]. But these cases require knowledge beyond the intrinsic fitness, which is a poor match for distributed network formation that assumes imperfect visibility of the network.

## 2.2 Central Limit Theorem

We intend to generate power law degree distributions without preferential attachment, which requires an alternative mechanism. A key insight is that mixtures of non-heavy-tailed random variables can produce heavy-tailed distributions. A generalized CLT shows that heavy tails tend to result from mixing geometric distributions with high variance parameters. This generalized CLT is distinct from the classical CLT familiarly used to justify statistics that assume normal distributions. The classical CLT states that the sample mean of independent random variables tends toward a normal distribution if the variables have finite mean and variance. A generalized CLT for variables with finite mean and infinite variance leads to heavy tails, even if the component distributions are not themselves heavy-tailed [10]. Over the limited range of a real network, distributions with high, but finite, variance may be approximated by this generalized CLT.  For example, a mixture of random variables with widely varying exponential distributions can approximate a power law [11].  The geometric distribution is the analog of the exponential distribution for discrete values, and a mixture of geometric distributions can also describe heavy-tailed distributions.  In fact, a mixture of geometric distributions with parameters pulled from a uniform distribution have high variance and therefore sum to a heavy-tailed distribution [12].  We use these insights both to create an alternative model for scale free network generation and to show that preferential attachment is only one of potentially many mechanisms that could lead to scale free networks.

## 3   Randomly stopped processes lead to power law distributions

Instead of this node growth and preferential attachment linking, our model considers that nodes begin with one link, and each link added is a discrete decision made in series.  After the first link, there is a chance the process stops.  If not, the node adds a second link and there is a chance to stop after that, and so on.  This stopping probability is the complement of the node fitness proposed in [8], ie the lower the fitness, the higher the stopping probability.  Node fitness in the modified BA model reflects the idea that nodes in real-world networks compete for links based on some intrinsic quality in addition to the extrinsic quality of degree that control preferential attachment.  In our model, we take the complement of this intrinsic fitness as our stopping probability but drop the preferential attachment from the modified BA model.  We represent this stopping probability with the intrinsic node property *q*.

The iterative, randomly stopped process is a plausible hypothesis for generating real-world networks.  Our randomly stopped linking model states that a webpage author will add one hyperlink, then decide whether stop.  If they continue, they add a third hyperlink before deciding whether to stop, and so on.  Each webpage has an intrinsic stopping probability reflecting how likely the author is to add links.  Stopping may be less likely for news articles with many hyperlinks but more likely for company advertising pages designed to funnel readers toward a particular action.  Undirected graphs have one stopping probability, while directed graphs can have two for the stochastic processes that determine each of the number of outgoing and incoming links.

**Theorem 1.** *Let there be a graph where each node has a parameter* q *drawn from a distribution uniform between* 0 *and* 1, *and has a degree* k *defined by:*

$$Pr(K = k) = q(1-q)^{k-1} \text{ where } k \in \{1, 2, 3, ...\}$$

Equation 3

*Let* p(k) *be the probability mass function (PMF) for node degrees in the graph.  Then, as the number of nodes and* k *are both large,* p(k) ∝ k$^{-2}$.

This theorem shows that scale free networks can form even when each individual node's degree is governed by a short-tailed distribution and links form randomly without preferential attachment.

Theorem 1 assumes a constant marginal probability to stop after each link, represented by the *shifted geometric distribution* (Equation 3): the number $K$ Bernoulli trials before the first success when each trial has success probability $q$. In our model, links are added after each failure and stopping occurs with the first successful trial. A single geometric distribution is not heavy-tailed and does not fit the high variance of a scale free network well. But if each node has a different stopping probability, the variance of parameter $q$ over all nodes can also be high. As pointed out in Section 2.2, mixing geometric distributions that have uniformly-distributed parameters results in a heavy-tailed distribution. Specifically, Theorem 1 assigns all nodes a link-stopping probability $q$ pulled randomly from a uniform distribution between *0* and *1*. Since the number of nodes is large, we represent the aggregated effect across nodes as an integration over all parameters $q$:

$$\int_0^1 q(1-q)^{k-1} \, dq = \frac{1}{k(k+1)} \text{ for } k > 0$$

Equation 4

This result is already a PMF without normalization

$$\sum_{k=1}^{\infty} \frac{1}{k^2 + k} = 1$$

Equation 5

and can therefore serve as a degree distribution: the probability to observe a node with degree $k$.

$$p(k) = \frac{1}{k^2 + k}.$$

Equation 6

We further approximate using a power law with $\gamma = 2$ because Equation 6 becomes $p(k) = k^{-2}$ as $k$ becomes large, which is consistent with the CLT described above and proves Theorem 1.

This section showed that a process using randomly stopped linking with widely varying stopping probabilities produces distributions following a power law. The next section shows how non-uniform mixing functions can achieve arbitrary values of $\gamma$.

### 3.1 Varying the mixing function to vary $\gamma$

In real networks, the scale free regime falls between $2 \leq \gamma \leq 3$ [13]. We showed that a uniform mixing coefficient for geometric distributions produces the degree distribution in Equation 6 which is approximated by a power law distribution with $\gamma = 2$. In comparison, the BA model has the degree distribution shown in Equation 2 and is approximated by a power law with $\gamma = 3$. To describe real networks, our model should be able to lead to arbitrary power law degree distributions with $2 \leq \gamma \leq 3$.

Equation 4 describes a mixture of geometric distributions with uniformly distributed parameters. To generalize, we can weight these distributions according to a mixing function *a(q)*. This represents the diversity of intrinsic stopping parameters for different nodes.

$$\int_0^1 a(q)q(1-q)^{k-1}\,dq$$
Equation 7

If properly normalized, *a(q)* is the PMF of intrinsic stopping parameters. We can see that Equation 4 is a special case of Equation 7 where *a(q)* = 1.

### 3.1.1 Linear Mixing Function

**Theorem 2.** *Let there be a graph like that in Theorem 1 except each node has a parameter q drawn from a linear distribution a(q) = 2q between 0 ≤ q ≤ 1. Then, as the number of nodes and k are both large, p(k) ∝ k⁻³.*

When the distribution of the random stopping parameter among nodes is *a(q) = 2q*, repeating the process for proving Theorem 1 yields:

$$p(k, a=2q) = \int_0^1 2qq(1-q)^{k-1}\,dq = \int_0^1 2q^2(1-q)^{k-1}\,dq = \frac{4}{k(k+1)(k+2)} \text{ for } k > 0$$
Equation 8

The limit of Equation 8 as *k* increases is a power law with γ = 3. In effect, the linear mixing function oversamples nodes with low stopping probability and under samples nodes with higher stopping probability compared to the uniform mixture. This results in a steeper power law than the uniform mixing case. The resulting mixture produces the same power law exponent as the BA model degree distribution in Equation 2.

### 3.1.2 Sublinear Mixing Function

Consider *a(q) = (1+c)q^c* where 0 ≤ *c* ≤ 1 and 0 ≤ *q* ≤ 1:

$$p(k, a=(1+c)q^c) = \int_0^1 (1+c)q^{1+c}(1-q)^{k-1}\,dq = \frac{(c+1)\Gamma(k)\Gamma(c+2)}{\Gamma(k+c+2)} \text{ for } k > 0$$
Equation 9

The right hand side of Equation 9 features the gamma function, which is an extension of the factorial function to complex numbers. For a positive integer,

$$\Gamma(k) = (n-1)!$$
Equation 10

Because *k* is constrained to positive integers, Equation 9 reduces to Equation 4 in the uniform mixing case when *c* = 0.

When $c = 0$, $\dfrac{(c+1)\Gamma(k)\Gamma(c+2)}{\Gamma(k+c+2)} \to \dfrac{\Gamma(k)\Gamma(2)}{\Gamma(k+2)} = \dfrac{(k-1)!1!}{(k+1)!} = \dfrac{1}{k(k+1)}$ for $k > 0$

Equation 11

Similarly, Equation 9 reduces to Equation 8 in the linear mixing case when *c = 1*.

When $c = 1$, $\dfrac{(c+1)\Gamma(k)\Gamma(c+2)}{\Gamma(k+c+2)} \to \dfrac{2\Gamma(k)\Gamma(3)}{\Gamma(k+3)} = \dfrac{2(k-1)!2!}{(k+2)!} = \dfrac{4}{k(k+1)(k+2)}$ for $k > 0$

Equation 12

When *0 ≤ c ≤ 1* and *k* grows large, Equation 9 approximates a power law with *2 ≤ γ ≤ 3*. Therefore, varying the mixing function varies *γ* to achieve the full range of power laws that arise from real-world scale free networks.

*a(q) = (1+c)q$^c$* is only one of infinitely many parameterizations of the mixing coefficient function in the randomly-stopped linking model. But it has the advantage of only requiring one parameter, *c*, to achieve any power law where *2 ≤ γ ≤ 3*, which is the entire range of scale free networks. We do not claim this mixing function is particularly likely to appear in a real-world network, but rather it shows that the mixing function does not require careful selection to produce degree distributions similar to real-world scale free networks.

This section demonstrated that the randomly stopped linking model, with the appropriate selection of intrinsic stopping probability values, can approximate an arbitrary scale free network's degree distribution. The next section describes a technique to generate synthetic networks according to the randomly stopped linking model.

## 3.2 Reparameterization of the Configuration Model to Generate Synthetic Networks

So far, we have analytically found the degree distribution of the randomly stopped linking model (Equation 9), but have not shown how to form a synthetic network by connecting nodes to other nodes with nondirectional links. We determine these links following the configuration model (CM) [14]:
1. Create a node.
2. Allocate a number of link stubs to this node.
3. Repeat 1 and 2 until all nodes are created.
4. Select a pair of nodes at random and create a link between them.
5. Repeat 4 until all stubs are linked.

This algorithm sets the degree distribution before links are formed. Therefore, degree distribution is determined only by intrinsic parameters of the nodes, with no network effects like preferential attachment. A scale free network created using this version of the CM is necessarily created without preferential attachment.

Our randomly stopped linking model is distinguished from other versions of the CM by how the stubs are allocated to nodes in step 2 above [15]. We pull the degree from a mixture of geometric distributions with high variance. To determine the number of link stubs for each node:
    2a. Assign the node a stopping probability from a random variable $q$ distributed between 0 and 1. This stopping probability is inversely proportional to the fitness.
    2b. Determine the number of link stubs according to a geometric distribution with $q$ as the parameter.
    2c. Connect these link stubs by selecting random pairs according to the CM until there are no more links remaining.

Figure 1 displays the results of this method as the degree distribution of four networks. The BA model generated one network by adding 2 links with each new node, connecting to existing nodes with probability proportional to the degree of the existing nodes. The other three networks were created by the randomly stopped linking model using stopping probability exponents $c=0$, $c=0.5$, and $c=1$. These stopping probabilities correspond to uniform, sublinear, and linear mixing distributions, respectively. Each is fit with a power law using the technique described in [16]. These results show that each method produces a power law with an exponent close to those expected.

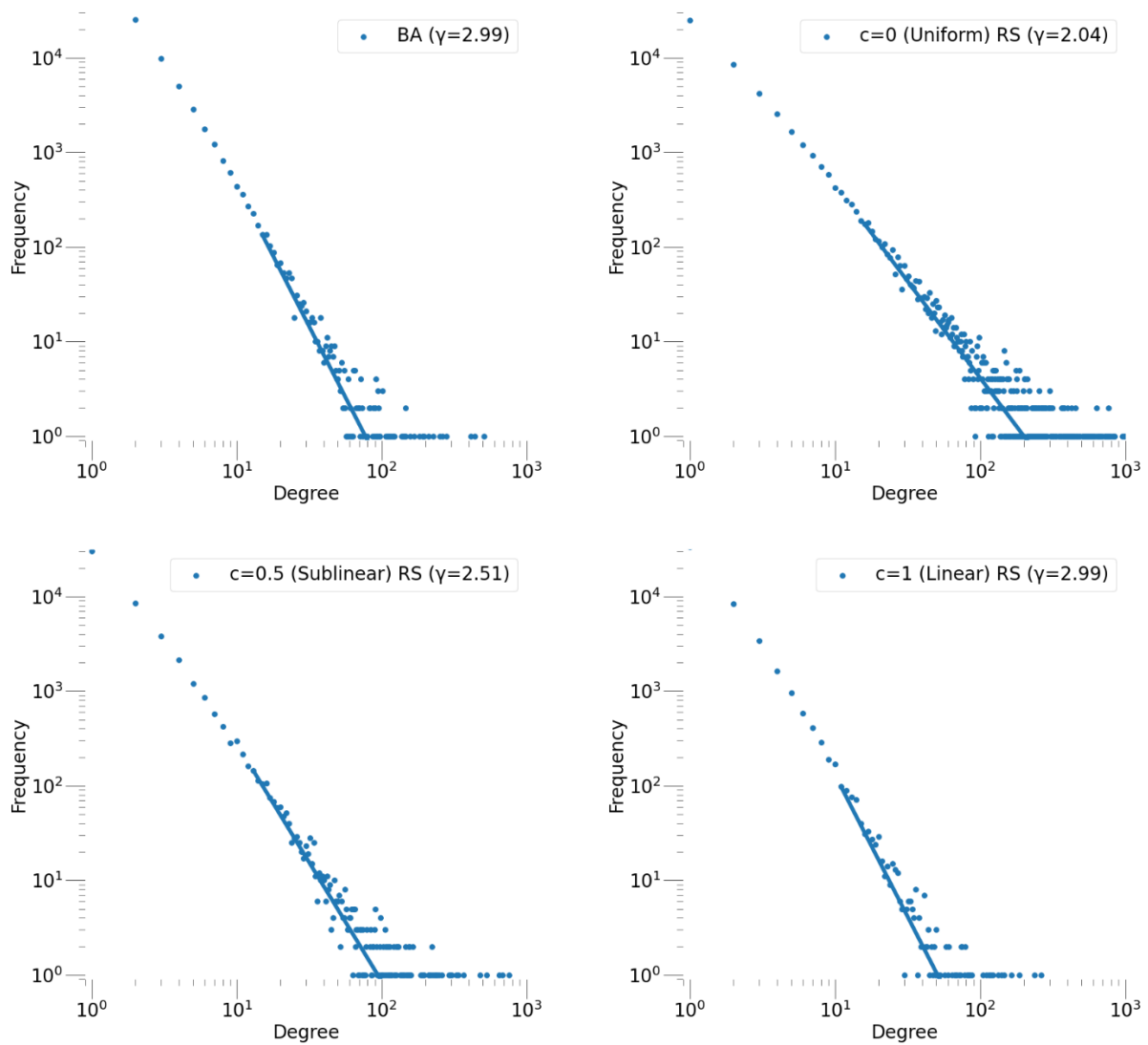

Figure 1 Networks generated using the BA model and the randomly stopped linking model each with 50,000 nodes.

Showing that our randomly stopped linking model can generate power law degree distributions doesn't prove that it is an important mechanism behind real-world scale free networks. But it does provide a plausible alternative that disproves preferential attachment as a necessary mechanism to generate scale free networks. The next section generalizes both the BA model and randomly stopped linking model to identify widely varying mixtures of geometric distributions as the common characteristic of mechanisms that lead to scale free networks.

## 4   Barabási Albert Model as a Bernoulli Process

There are obvious similarities between the degree distribution of the BA model (Equation 2) and the degree distribution of the randomly stopped linking model with linear mixing coefficient (Equation 8). In this section, we show that is because the BA model draws random variables from a collection of geometric distributions with high variance and therefore, according to the CLT described above, tends toward a power law. The high variance in this repeated Bernoulli process, not preferential attachment per se, is why the BA model generates power law degree distributions.

Similarly to [7], we view the BA model as a process where a link is added at each timestep *t* and nodes are added at the integer values of *t/m*. (Remember that in the BA model, *m* is the number of links added with each new node and is constant for a given network.) At each timestep, a link is added to connect from the newest node to some node in the graph. As in the approach in [7], we allow self-links or duplicates to links that already exist. The original formulation of the BA model is ambiguous on this point and these cases are in any event rare in large networks. The probability of a given node being the target of this new link is given by Equation 1, which we can rewrite in terms of *t*:

$$\Pi(k_i, t) = \frac{k_i}{2t}$$

Equation 13

The BA model is a Bernoulli process with a trial at each timestep t for each node in the graph because there are two possible outcomes: a link is added or it is not. The probability of a link being added at each timestep for each node is given by Equation 13 in terms of each node's degree.

The linking events are Bernoulli trials pulled from a set of distributions that are unique to each node and timestep combination. However, the problem can be simplified since all nodes with the same degree at the same timestep share the same preferential attachment coefficient and therefore the same probability of receiving a new link. Given that the maximum degree in real networks is usually only a few hundred, there are relatively few distinct probabilities at any given timestep. What's more, the space of possible changes in the population of nodes sharing a degree is small. There are two ways for a node to have degree *k* at timestep *t+1*:
   1. a node with degree *k* fails to receive the new link formed at timestep *t*
   2. a node with degree *k-1* receives the new link formed at timestep *t*

At timestep *t*, the expected fraction of nodes with degree *k* is given by Equation 2. We find the expected fraction of nodes with degree *k* at timestep *t+1* by aggregating the Bernoulli trials across all nodes sharing degree *k* or *k-1* to determine the likelihood of the cases above. This results in the pmf at timestep *t+1*:

$$\frac{\frac{t}{m}}{\frac{t+1}{m}} \left( (1 - \frac{k}{2t}) \cdot \frac{2m(m+1)}{k(k+1)(k+2)} + \frac{k-1}{2t} \cdot \frac{2m(m+1)}{(k-1)k(k+1)} \right)$$

Equation 14

At timestep t, the probability of failure in the Bernoulli trial for each node with degree k is given by the complement of Equation 13
Equation 13

where $k_i$ is set to $k$.  Since all nodes sharing the same degree share the same linking probability, we weight this trial by the fraction of nodes with degree $k$, which is given by Equation 2.  This yields the first term of Equation 14:

$$(1 - \frac{k}{2t}) \cdot \frac{2m(m+1)}{k(k+1)(k+2)}$$
Equation 15

Next, we'll consider the second case, where a node with degree k-1 receives an additional link. At timestep t, the probability of success in the Bernoulli trial for each node with degree k-1 is given by Equation 13 where $k_i$ is set to *k-1*.  The fraction of nodes with degree *k-1* is found by replacing *k* with *k-1* in Equation 2.  Weight this trial yields the second term of Equation 14:

$$\frac{k-1}{2t} \cdot \frac{2m(m+1)}{(k-1)k(k+1)}$$
Equation 16

Finally, we account for the increase in graph size by multiplying these terms by the ratio of number of nodes at timestep *t* to the number of nodes at timestep *t+1*. These Bernoulli trials occur at each link creation and there are *m* links per new node, so the number of nodes in the graph is *t/m*.  This value can be non-integral, reflecting that a new node is not fully added to the graph until it has created *m* links.

Putting these terms together results in Equation 14, which reduces to exactly Equation 2, demonstrating the validity of the interpretation of the BA model as a mixture of Bernoulli processes.

That the BA model can be represented by Bernoulli processes is insufficient to explain a power law degree distribution.  It also needs a widely varying mixture of success probabilities.  In the BA model, a node's linking probability changes with each added link because the value of t in the denominator of Equation 13 increments.  Even when the Bernoulli trial succeeds and the numerator of Equation 13 increments, it still increases by a factor of two less than the denominator.  This means trials run early in the graph when *t* is small have much higher success probabilities than those run later.  Change in the preferential attachment coefficient over time is therefore the driver that creates high variance even though many nodes may share the same linking probability at any given *t*.  This contrasts with the randomly stopped linking model, where each node has a constant linking probability over time but there is high variance between nodes.

The similarity does not imply the BA model and randomly stopped model with intrinsic stopping probability are the same process.  The BA model uses preferential attachment and growth, forming links based on the degree of the nodes in the existing network up to the point in time a new node is added.  The randomly stopped model uses a temporally constant stopping probability that is an intrinsic property of each node and does not depend on degree or any other network structure.  Networks created by each model may have significantly different structures even with the same degree distribution.

The similar degree distributions arise because both models run a series of Bernoulli trials with widely varying probabilities that can be represented by a mixture of distributions with widely varying parameters. The CLT shows that in these circumstances, a power law is expected, not anomalous. Therefore, preferential attachment and growth in the BA model are properly understood as just one among many configurations of Bernoulli trials that result in power law degree distributions, rather than a necessary component of the processes that lead to scale free networks.

## 5   Conclusions and Future Work

We demonstrate a randomly stopped linking model that could arise from plausible physical processes in the real-world and generates scale free networks when used to reparametrize the CM. A CLT suggests power law degree distributions are expected when samples are taken from Bernoulli trials where the probabilities have high variance. Finally, we reinterpret the BA model to show that preferential attachment is itself a series of Bernoulli trials following probabilities with high variance.

Preferential attachment is not a requirement for scale free networks. Any physical process represented as Bernoulli trials with high variance should generate them. Much of the network science literature assumes the BA model's preferential attachment mechanism must be a dominant process in real-world networks because it has long been the only robust explanation of scale free network formation. The existence of an alternative explanation in the randomly stopped linking model shows that preferential attachment cannot be assumed to be the primary – or even a significant – mechanism in real-world networks. Further, the generalized CLT for Bernoulli trials with high variance makes it unlikely that preferential attachment and randomly stopped linking are the only mechanisms that lead to scale free networks.

In the real-world, with its great variety of outcomes and many repeated trials, scale free networks are exactly what we should expect. Exponential networks of the Erdos-Renyi model are the unnatural exception, encountered only when outcomes are narrowly banded within a single outcome probability.

Realizing that preferential attachment is not the only mechanism to generate scale free networks, future work in applied network science should seek to identify mechanisms in specific, real-world networks that follow the general pattern of Bernoulli trials with high variance. Also, this research focused on degree distribution, which is only one important network characteristic. Additional work should consider how expectations for network diameter, component size, average path length, neighborhood clustering, and other network parameters differ between the BA model, the randomly stopped linking model, and other mechanisms yet to be identified. Comparing those model-generated expectations to the parameters of real-world networks may measure the relative contribution of different mechanisms in real-world systems.

# 6 References

[1] Barabási, A.-L. Scale-Free Networks: A Decade and Beyond. *Science* **325,** 412–413 (2009).

[2] Barabási, A.-L. Emergence of Scaling in Random Networks. *Science* **286,** 509–512 (1999).

[3] Albert, R., Jeong, H. & Barabási, A.-L. Diameter of the World-Wide Web. *Nature* **401,** 130–131 (1999).

[4] Broido, A. D. & Clauset, A. Scale-free networks are rare. *Nature Communications* **10,** (2019).

[5] Holme, P. Rare and everywhere: Perspectives on scale-free networks. *Nature Communications* **10,** (2019).

[6] Stumpf, M. P. H. & Porter, M. A. Critical Truths About Power Laws. *Science* **335,** 665–666 (2012).

[7] Bollobás, B., Riordan, O., Spencer, J. and Tusnády, G. (2001), The degree sequence of a scale-free random graph process. Random Struct. Alg., 18: 279-290. https://doi.org/10.1002/rsa.1009

[8] Bianconi, G., Barabási, A.-L. Competition and Multiscaling in Evolving Networks. *Europhysics Letters (EPL),* 54(4):436-442, 2001.

[9] Servedio, V. D. P., Caldarelli, G., Buttà, P. Vertex Intrinsic Fitness: How to Produce Arbitrary Scale-free Networks. *Physical Review E,* 70(5), 2004.

[10] Willinger, W., Alderson, D., Doyle, J. & Li, L. More Normal Than Normal: Scaling Distributions and Complex Systems. *Proceedings of the 2004 Winter Simulation Conference, 2004.* doi:10.1109/wsc.2004.1371310

[11] Chen, J. From the central limit theorem to heavy-tailed distributions. *Journal of Applied Probability* **40**, 803–806 (2003).

[12] Balanzario, E. P., Ramírez, R. M. & Ortiz, J. S. The randomly stopped geometric Brownian motion. *Statistics & Probability Letters* **90,** 85–92 (2014).

[13] Barabási, A.-L., Pósfai, M. (2016). *Network science*. Cambridge: Cambridge University Press. ISBN: 9781107076266 1107076269

[14] Newman, M. E. J. The Structure and Function of Complex Networks. *SIAM Review*, 45(2):167-256, 2003.

[15] Johnston, J. & Andersen, T. Randomly Stopped Linking Generates Scale Free Networks. *SIAM Workshop on Network Science 2020*, Toronto (July 9-10, 2020)

[16] Clauset, A., Shalizi, C. R., and Newman, M. E. J. Power-law Distribution in Empirical Data. *SIAM Reivew*, 514(4);661-703, April 2009.